\documentclass{article}
\PassOptionsToPackage{numbers, compress}{natbib}
\usepackage[preprint]{neurips_2020}

\usepackage[utf8]{inputenc} % allow utf-8 input
\usepackage[T1]{fontenc}    % use 8-bit T1 fonts
\usepackage{hyperref}       % hyperlinks
\usepackage{url}            % simple URL typesetting
\usepackage{booktabs}       % professional-quality tables
\usepackage{amsfonts}       % blackboard math symbols
\usepackage{nicefrac}       % compact symbols for 1/2, etc.
\usepackage{microtype}      % microtypography
\usepackage{hyperref}

\usepackage{graphicx}

\usepackage{xcolor, soul}
\definecolor{orange}{rgb}{1,0.5,0}
\definecolor{mydarkcyan}{rgb}{0,0.5,0.5}

\title{Astronomical Image Quality Prediction based on Environmental and Telescope Operating Conditions}

\author{Sankalp Gilda \\
    University of Florida \\
    Gainseville, FL 32611, United States \\
    \texttt{s.gilda@ufl.edu} \\
  \And
  Yuan-Sen Ting\\
  Institute for Advanced Study\\
  Princeton, NJ 08540, United States\\
   \texttt{ting@ias.edu} \\
  \And
  Kanoa Withington\\
   Canada-France-Hawaii Telescope \\
 Kamuela, HI, United States \\
  \texttt{kanoa@cfht.hawaii.edu} \\
 \And
 Matt Wilson\\
  Canada-France-Hawaii Telescope \\
Kamuela, HI, United States \\
 \texttt{wilson@cfht.hawaii.edu} \\
\And
Simon Prunet\\
Canada-France-Hawaii Telescope \\
Kamuela, HI, United States \\
 \texttt{prunet@cfht.hawaii.edu} \\
\And
  William Mahoney\\
    Canada-France-Hawaii Telescope \\
  Kamuela, HI, United States \\
   \texttt{billy@cfht.hawaii.edu} \\
  \And
  Sebastien Fabbro\\
  National Research Council Herzberg \\
  4071 West Saanich Road, Victoria, BC, Canada \\
   \texttt{Sebastien.Fabbro@nrc-cnrc.gc.ca} \\
  \And
  Stark C. Draper \\
  University of Toronto\\
  Toronto, ON M5S 3G4, Canada \\
   \texttt{stark.draper@utoronto.ca} \\
  \And 
  Andy Sheinis\\
  Canada-France-Hawaii Telescope \\
  Kamuela, HI, United States \\
   \texttt{sheinis@cfht.hawaii.edu} \\
}

\begin{document}

\maketitle

\begin{abstract}
    Intelligent scheduling of the sequence of scientific exposures taken at ground-based astronomical observatories is massively challenging.  Observing time is over-subscribed and atmospheric conditions are constantly changing. We propose to guide observatory scheduling using machine learning. Leveraging a 15-year archive of exposures, environmental, and operating conditions logged by the Canada-France-Hawaii Telescope, we construct a probabilistic data-driven model that accurately predicts image quality.  We demonstrate that, by optimizing the opening and closing of twelve vents placed on the dome of the telescope, we can reduce dome-induced turbulence and improve telescope image quality by (0.05-0.2 arc-seconds).  This translates to a reduction in exposure time (and hence cost) of $\sim 10-15\%$. Our study is the first step toward data-based optimization of the multi-million dollar operations of current and next-generation telescopes. 
\end{abstract}

\section{Introduction}\label{sec:introduction} 
We describe initial steps towards the first-ever machine-learning driven approach for real-time scheduling of astronomical observations at the Canada-France-Hawaii Telescope (CFHT). Situated at the summit of the 4,200m volcano of Mauna Kea on the island of Hawai'i, CFHT is one of the world's most productive ground-based observatories~\citep{crabtree:2019}.  CFHT has gathered decades-worth of observational data since it saw first light in 1979 \citep{racine1991mirror}. We train a deep neural network on this  data archive to predict the ``image quality'' (IQ) of each candidate scientific exposure as a function of environmental and observatory dome operating parameters. Our long-term goal is to leverage such predictors to schedule observations and dome controls to maximize IQ. Real-time scheduling is key since (i) observing time at CFHT is oversubscribed with science proposals by $\sim 3$ fold, and (ii) CFHT costs around USD 25,000 to operate per night (a tenth of what next generation 30m-class telescopes will cost). Improving the efficiency of operations will yield more science, and will pave the way for the more complicated scheduling tasks of next-generation telescopes.

\begin{figure}
   \centering
   \includegraphics[width=0.8\linewidth]{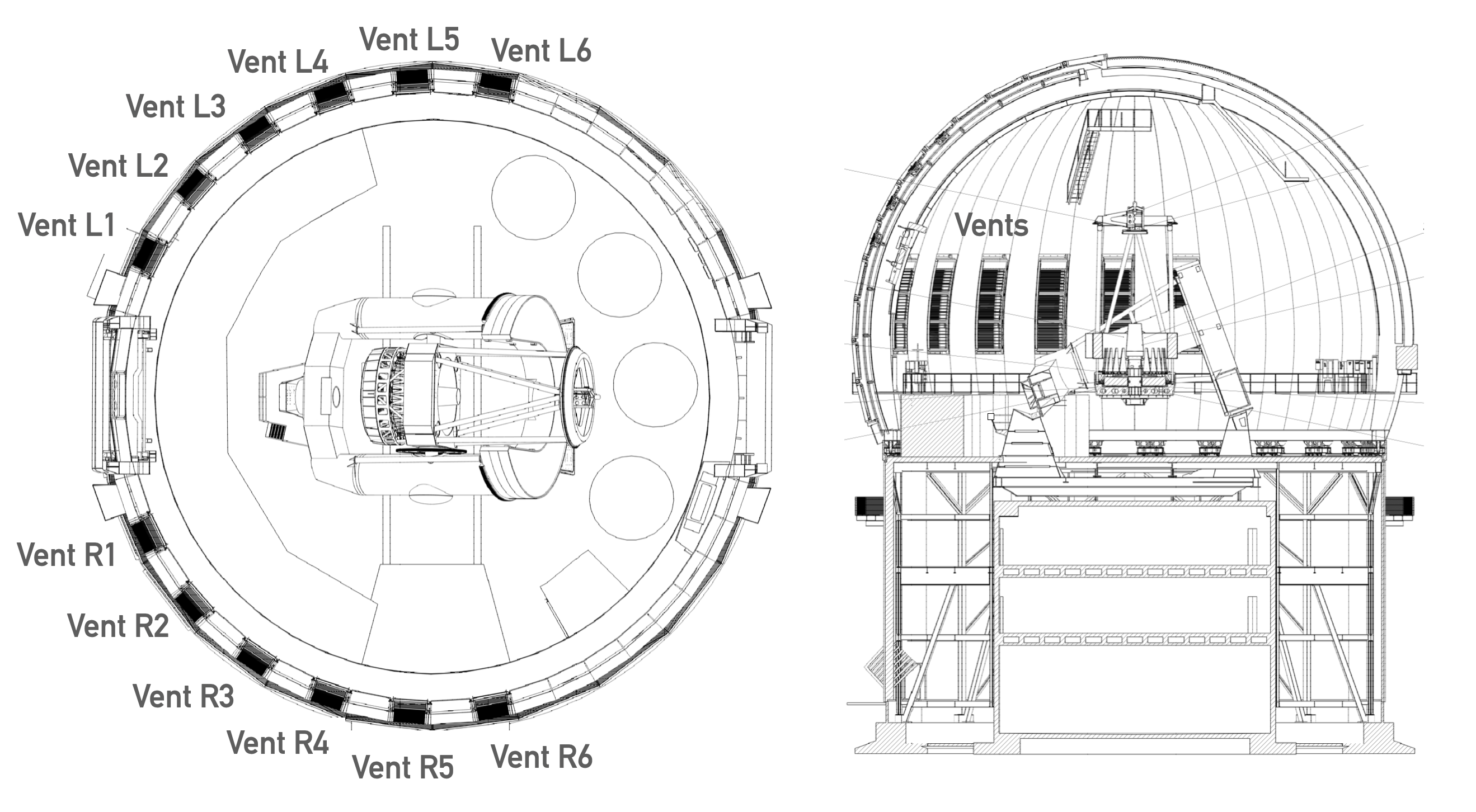}
   \caption{A schematic of CFHT. The twelve actuable dome vents are marked.}
   \label{fig:dome}
\end{figure}

IQ measures the point-source blurring of a star and relates directly to signal-to-noise ratio (SNR). IQ degradation from the theoretical maximum results from turbulence both in the atmosphere and from the night-time cooling of the observatory dome. The summit of Mauna Kea is a prime site for an observatory as atmospheric turbulence is minimal due to the smooth flow of the prevailing trade winds and the height of the summit.  However, despite continual improvements in CFHT IQ since 1979, including the 2012 introduction of vents to assist in flushing hot air from the dome (see Fig.~\ref{fig:dome}), as with all major ground-based observatories, IQ rarely reaches what the site can theoretically deliver.

Through the implementation of a probabilistic DNN, we demonstrate that  ancillary environmental and operating parameter data are sufficient to predict IQ accurately. We illustrate that, keeping all other setting constant, there exist optimal configurations of dome vents that can substantially improve IQ. Our successes here lay the foundation for developing automated control and scheduling approaches.

\section{Data} \label{sec:data}

Almost two decades worth of MegaPrime (the CFHT wide field optical instrument) IQ measurements have been archived at CFHT \cite{megacam}. Prior to their use in our network, we needed to clean the data substantially: pertinent variables were spread across multiple data sets, records were missing measurements due to sensor failures, and contained errant variable values due to mis-calibrated data reduction pipelines. We have made the cleaned data publicly available.\footnote{
\texttt{http://www.cfht.hawaii.edu/}. {\it We clipped the URL in order to respect the double-blind review process.}}

Each input record contains three distinct types of predictive variables: (1) {\em Dome operating parameters:} These include the configurations of the twelve dome vents (open or closed), and the windscreen setting (degrees open). These are examples of the variables that we can adjust and optimize the settings of in real time. In this initial  study we focus on the dome vents. (2) {\em Environmental parameters:} These include  exposure-averaged wind speed, wind direction, barometric pressure, and temperature values at various points both inside and outside the observatory.
(3) {\em Ancillary parameters:} These include, for each observation, the telescope altitude and azimuth (which correspond to the astronomical object being observed), the color filter, and the length of exposure. In total, there are 86 variables (including IQ) provided with each exposure. We engineer these as follows: the vent status are binarized as 0 and 1, and every temperature variable is subtracted from every other temperature variable to enable ease capture of secondary effects. Variables with $>50\%$ missing or errant values are dropped, followed by samples with any missing values. The final dataset thus obtained has 1,247 variables and 69,846 records, spanning observations made from May 2005 to March 2020.

\section{Method} \label{sec:methods}

\begin{figure}
   \centering
   \includegraphics[width=1.0\linewidth]{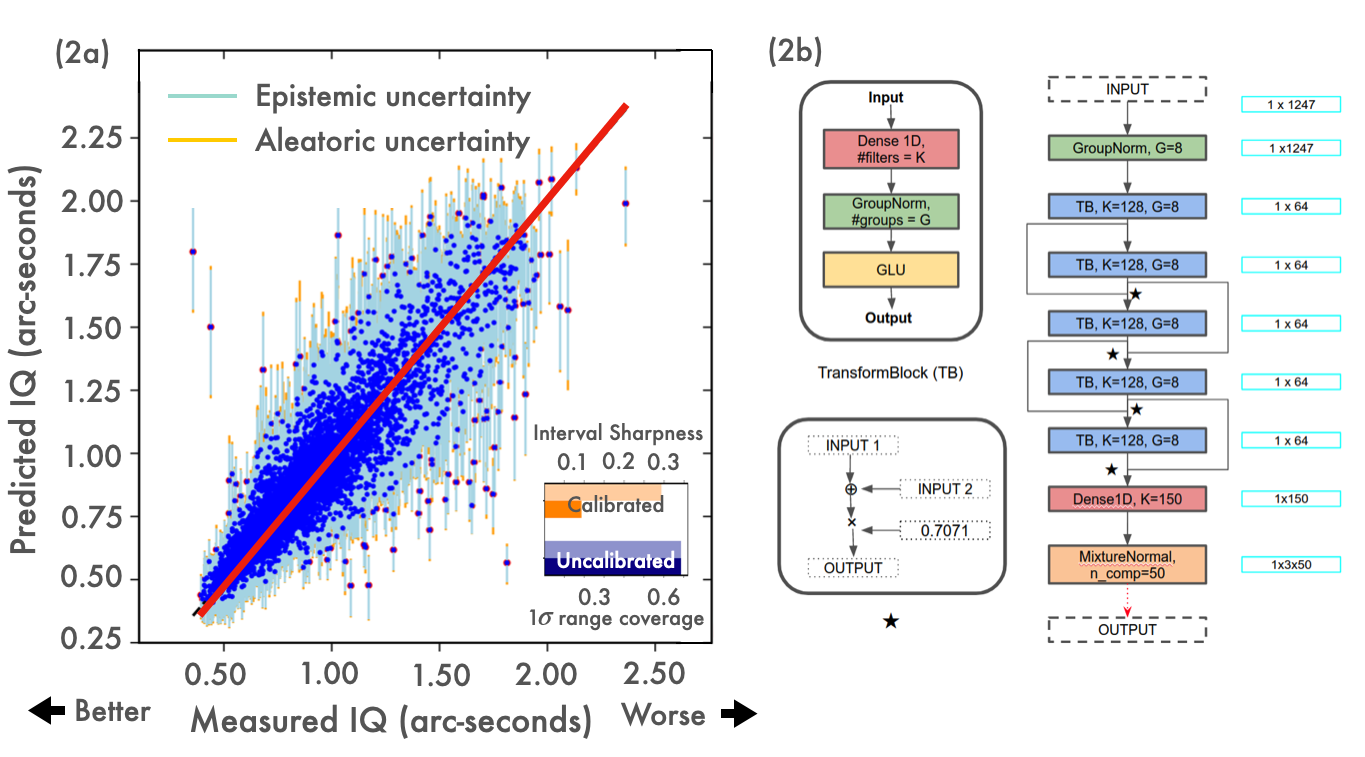}
   \caption{ \textbf{(2a):} Telescope IQ can be predicted to, on average, 0.07 arc-sec based on configurable telescope parameters (e.g., vent configuration) and  environmental conditions (e.g., wind speed, bolometric pressure). The inset shows the "sharpness" of the prediction and the range covered by the predicted one-sigma uncertainty.  This demonstrates that the CRUDE calibration improves the robustness of the probabilistic outputs. \textbf{(2b):} The network architecture used in this study: a dense feed-forward network with skip connections and GLU activations.}
   \label{fig:1-to-1_plus_mdn}
\end{figure}

We implement a feed-forward neural network (see Fig. (\ref{fig:1-to-1_plus_mdn}b)) with 5 dense layers, 64 neurons per layer, skip connections, and generalized linear units \citep{tabnet}. We wrap the network as part of a mixture density network (MDN) \cite{bishop_mdn} to provide probabilistic output of the image quality, comprised of a mixture of a weighted sum of 50 Beta distributions. The network is optimized using adaptive gradient descent and the negative log likelihood loss function.

The nature of the MDN allows us to predict a PDF of IQ for each sample. For data sample $i$ and mixture model component $j$, let $\mu_{i,j}$, $\sigma_{i,j}^2$, and $\omega_{i,j}$ respectively denote the mean, variance, and normalized weight in the mixture model. We obtain the predicted IQ value as the weighted mean of the individual modes, $\mu_i = \sum_{j=0}^{m-1} \omega_{i,j} \mu_{i,j}$. Aleatoric uncertainty (or ``irreducible uncertainty'') \citep{gal_thesis, kendall2017uncertainties} quantifies the inherent stochasticity of the inputs. We use a weighted average of the mixture model variances $\sigma_{i,a}^2 =  \sum_{j=0}^{m-1} \omega_{i,j} \sigma_{i,j}^{2}$ as its estimator \citep{choi2018uncertainty}. Epistemic uncertainty (or ``reducible uncertainty'') \citep{gal_thesis, kendall2017uncertainties} quantifies the model's uncertainty in parts of the parameter space sparsely populated by the training samples. We use the weighed variance of the mixture model means, $\sigma_{i,e}^2 = \sum_{j=0}^{m-1} \omega_{i,j} \mu_{i,j}^{2} - \mu_i^{2}$ \citep{choi2018uncertainty}. 
Total uncertainty is computed by adding these in quadrature.

Noting that, in practice, most uncertainty estimates are over-confident when estimating the prediction uncertainty \citep{lakshminarayanan_probability_calibration0, probability_calibration1, crude_probability_calibration},  we use the CRUDE method \cite{crude_probability_calibration} to re-calibrate our model's predictions \emph{post-hoc}. CRUDE ensures that not only are the post-processed predictions calibrated, but that they are also \emph{sharp} \citep{measuring_calibration_in_deep_learning}. Sharpness refers to the concentration of the predictions.  The more peaked (the sharper) the predictions are, the better (provided it is not at the expense of calibration).

We split the total input data into training and test sets in a ratio of 9-to-1. The training data is the subdivided into training and validation subsets via cross-validation in a ratio of 9-to-1. The model predictions based on the validation splits are concatenated and used to calibrate the predictions on the held-out test set.

\section{Results} 
\label{sec:results}

In Fig. (\ref{fig:1-to-1_plus_mdn}a), we compare the predicted IQ as a function of the IQ measured on the astronomical images. Error bars show both the epistemic and aleatoric parts of the prediction error obtained by the MDN. The figure shows that IQ can be predicted without bias and with reasonable accuracy using only measured environmental conditions and dome operating parameters as inputs. The mean absolute error is 0.07 arc-seconds (or arc-sec, for simplicity). The inset demonstrates the benefit of \emph{post-hoc} calibration with CRUDE: for samples in our test set, the predicted 1-$\sigma$ uncertainties much better capture the expected 68\% of the true IQ values, and are also small enough so that the resulting probability distribution functions are sharply peaked.

\begin{figure}
   \centering
    \includegraphics[width=\textwidth]{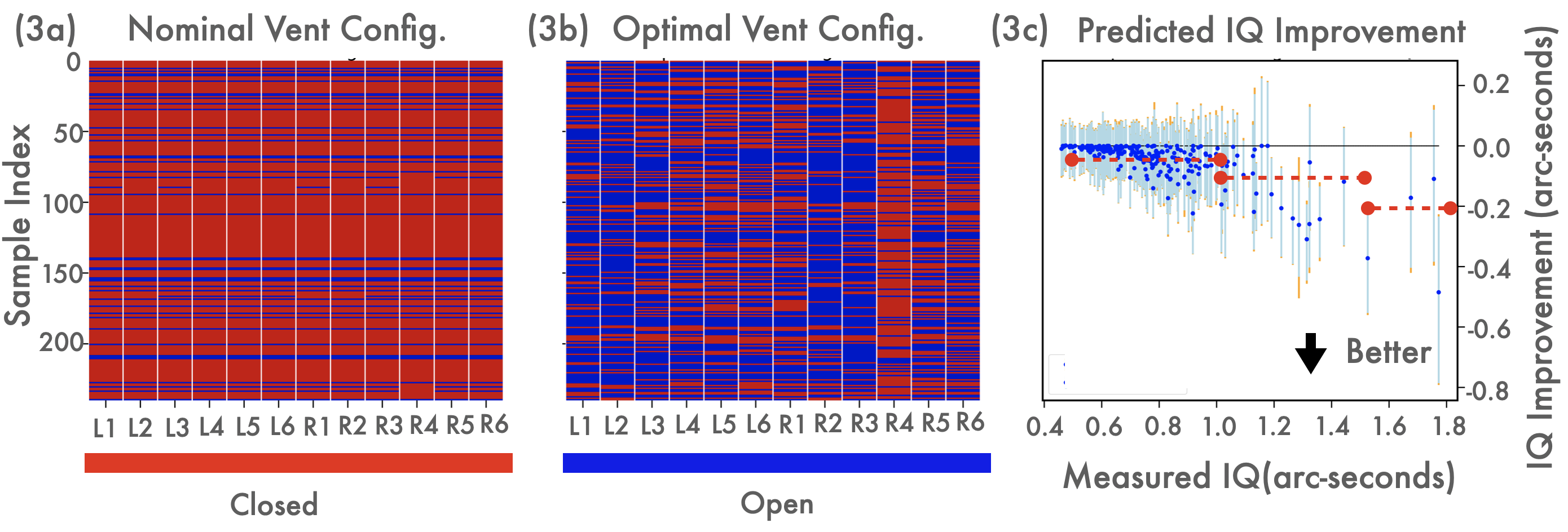}
    \caption{Optimizing the vent configuration can dramatically improve the image quality of the telescope. \textbf{(3a)} shows the nominal vent configurations, and \textbf{(3b)} the (predicted) optimal vent configuration that leads to the IQ improvements shown in \textbf{(3c)}. The red dashed lines in \textbf{(3c)} delineate the assumed improvements for the exposure time calculation.}
    \label{fig:actionable}
\end{figure}

As already mentioned, the dome vents (see Fig. \ref{fig:dome}) were installed in 2012 to flush hot interior dome air in an attempt to reduce dome-induced turbulence \citep{bauman2014dome, salmon2009cfht, Pfrommer_2015}. In Fig.~\ref{fig:actionable} we investigate the effect of these vents in some detail. By discovering relationships between vent configurations and environmental conditions we show that there exist optimal the vent configurations (Fig. (\ref{fig:actionable}b)) that could improve observed IQ (Fig. (\ref{fig:actionable}c)). Fig. (\ref{fig:actionable}a) visualizes the ``nominal'' vent configuration in effect when data was collected for a randomized sample of 250 exposures. The red/blue horizontal lines indicate the configuration of the 12 vents with red denoting closed and blue denoting open. The figure shows that vents have historically been mostly all closed or all open. However, our model can make predictions across all possible $2^{12}$ vent configurations. We can therefore ask which of the possible vent configurations would result in the best (the lowest) IQ.  The (predicted) IQ-minimizing vent configuration for each of the 250 exposures is shown in Fig. (\ref{fig:actionable}b). The difference between the nominal and optimal vent configurations is immediately apparent. The latter are often a mixture of open and close vents, something not currently implemented at CFHT.

The improvement (decrease) in predicted IQ is shown in Fig. (\ref{fig:actionable}c). When nominal IQ is excellent ($<$ 0.5 arc-sec), we see that the possible improvement in IQ is marginal. In this regime atmospheric turbulence, over which we have no control, dominates IQ. However, when IQ is middling or degrades ($>$ 0.5 arc-sec), by optimizing vent configuration we can substantially improve the (predicted) IQ. The improvement ranges from about 0.05 arc-sec for a nominal IQ of 0.5-1.0 arc-sec, to about 0.1 arc-sec for a nominal IQ of 1.0-1.5 arc-sec, and to 0.2 arc-sec for a nominal IQ $>$1.5 arc-sec. This range of IQ captures most observations; it is rare to have nominal IQ $<0.5$ arc-sec. Using the CFHT exposure time calculator\footnote{\texttt{http://etc.cfht.hawaii.edu/mp/}. We assumed point sources of 20-23 mag in the r-band filter.}, we find that these IQ improvements correspond to a $10-15\%$ reduction in exposure time (and hence cost). The result is largely independent of the nominal IQ, target SNR, and source brightness (as evidenced by the the red dashed lines in Fig. (\ref{fig:actionable}c)).

\section{Discussion and Future Directions} \label{sec:discussion}

This study applies a probabilistic network to fifteen years of data collected by the Canadian-France-Hawaii Telescope. We demonstrate that on average image quality can be predicted to 0.07 arc-sec accuracy based on environmental conditions and telescope operating parameters. By varying the configuration of the dome vents in response to extant environmental conditions, we show that IQ can be improved by about 10\% across all regimes, with larger gains when the nominal IQ value is large.  Given a target SNR of an astronomical observation, this translates to a reduction in observation time and costs of up to 10-15\% (about 1M USD a year). Overall, this study is an important first step towards active and automated optimization of telescope IQ, and eventual real-time scheduling.

We will pursue several avenues of further inquiry. The improved IQs we present here are predicted by extrapolating over hypothetical vent configurations.  While the uncertainties output by our model suggests that these marginal ``out-of-distribution'' predictions are most likely robust, they must be verified in real-life. We also treat the input data as independent and thus do not leverage various temporal relationships. Correlations between adjacent time stamps contains implicit information that we plan to exploit using specialized architectures such as LSTMs \citep{lstm} to predict IQ in real-time -- some 5 to 10s of minutes out -- to enable adaptive reorganization of the nominal observation schedule. Finally, in future work we will explore use of variational encoders for feature imputation \citep{vae_missingvalues}, since it is known that dropping samples negatively impacts a model's predictive ability \citep{imputation_missingvalues}.

\section*{Broader Impact}

This study leverages deep learning to exploit existing data catalogs at the Canada-France-Hawaii Telescope to improve telescope operations. Large ground-based observatories like CFHT are enormously expensive to build and operate, have limited observing time that is heavily over-subscribed with meritorious science proposals, and play a key role in furthering our understanding of the universe and of ourselves. By capitalizing on data that is already in hand, we can improve operations, gain novel insights, and develop techniques that can be used to advance the scientific agenda of observatories worldwide. We do not foresee any ethical concerns arising from this work.

{\small
\bibliographystyle{IEEEtran}%IEEEtran
\bibliography{bibliography,neurips}

\begin{thebibliography}{10}
\providecommand{\url}[1]{#1}
\csname url@samestyle\endcsname
\providecommand{\newblock}{\relax}
\providecommand{\bibinfo}[2]{#2}
\providecommand{\BIBentrySTDinterwordspacing}{\spaceskip=0pt\relax}
\providecommand{\BIBentryALTinterwordstretchfactor}{4}
\providecommand{\BIBentryALTinterwordspacing}{\spaceskip=\fontdimen2\font plus
\BIBentryALTinterwordstretchfactor\fontdimen3\font minus
  \fontdimen4\font\relax}
\providecommand{\BIBforeignlanguage}[2]{{%
\expandafter\ifx\csname l@#1\endcsname\relax
\typeout{** WARNING: IEEEtran.bst: No hyphenation pattern has been}%
\typeout{** loaded for the language `#1'. Using the pattern for}%
\typeout{** the default language instead.}%
\else
\language=\csname l@#1\endcsname
\fi
#2}}
\providecommand{\BIBdecl}{\relax}
\BIBdecl

\bibitem{crabtree:2019}
\BIBentryALTinterwordspacing
D.~Crabtree, ``{Canada's astronomy performance based on bibliometrics},'' Oct.
  2019, white paper identifier W014. [Online]. Available:
  \url{https://doi.org/10.5281/zenodo.3756124}
\BIBentrySTDinterwordspacing

\bibitem{racine1991mirror}
R.~Racine, D.~Salmon, D.~Cowley, and J.~Sovka, ``Mirror, dome, and natural
  seeing at cfht,'' \emph{Publications of the Astronomical Society of the
  Pacific}, vol. 103, no. 667, p. 1020, 1991.

\bibitem{megacam}
\BIBentryALTinterwordspacing
O.~Boulade, X.~Charlot, P.~Abbon, S.~Aune, P.~Borgeaud, P.-H. Carton, M.~Carty,
  J.~D. Costa, H.~Deschamps, D.~Desforge, D.~Eppelle, P.~Gallais, L.~Gosset,
  R.~Granelli, M.~Gros, J.~de~Kat, D.~Loiseau, J.~L. Ritou, J.~Y. Rousse,
  P.~Starzynski, N.~Vignal, and L.~G. Vigroux, ``{Megacam: the new
  Canada-France-Hawaii Telescope wide-field imaging camera},'' in
  \emph{Instrument Design and Performance for Optical/Infrared Ground-based
  Telescopes}, M.~Iye and A.~F.~M. Moorwood, Eds., vol. 4841, International
  Society for Optics and Photonics.\hskip 1em plus 0.5em minus 0.4em\relax
  SPIE, 2003, pp. 72 -- 81. [Online]. Available:
  \url{https://doi.org/10.1117/12.459890}
\BIBentrySTDinterwordspacing

\bibitem{tabnet}
S.~O. Arik and T.~Pfister, ``Tabnet: Attentive interpretable tabular
  learning,'' \emph{arXiv preprint arXiv:1908.07442}, 2019.

\bibitem{bishop_mdn}
C.~M. Bishop, \emph{Neural Networks for Pattern Recognition}.\hskip 1em plus
  0.5em minus 0.4em\relax USA: Oxford University Press, Inc., 1995.

\bibitem{gal_thesis}
Y.~Gal, ``Uncertainty in deep learning,'' in \emph{Uncertainty in deep
  learning}.\hskip 1em plus 0.5em minus 0.4em\relax University of Cambridge,
  2016.

\bibitem{kendall2017uncertainties}
A.~Kendall and Y.~Gal, ``What uncertainties do we need in bayesian deep
  learning for computer vision?'' in \emph{Advances in neural information
  processing systems}, 2017, pp. 5574--5584.

\bibitem{choi2018uncertainty}
S.~Choi, K.~Lee, S.~Lim, and S.~Oh, ``Uncertainty-aware learning from
  demonstration using mixture density networks with sampling-free variance
  modeling,'' in \emph{2018 IEEE International Conference on Robotics and
  Automation (ICRA)}.\hskip 1em plus 0.5em minus 0.4em\relax IEEE, 2018, pp.
  6915--6922.

\bibitem{lakshminarayanan_probability_calibration0}
B.~Lakshminarayanan, A.~Pritzel, and C.~Blundell, ``Simple and scalable
  predictive uncertainty estimation using deep ensembles,'' in \emph{Advances
  in neural information processing systems}, 2017, pp. 6402--6413.

\bibitem{probability_calibration1}
V.~Kuleshov, N.~Fenner, and S.~Ermon, ``Accurate uncertainties for deep
  learning using calibrated regression,'' \emph{arXiv preprint
  arXiv:1807.00263}, 2018.

\bibitem{crude_probability_calibration}
E.~Zelikman and C.~Healy, ``Improving regression uncertainty estimates with an
  empirical prior,'' \emph{arXiv preprint arXiv:2005.12496}, 2020.

\bibitem{measuring_calibration_in_deep_learning}
J.~Nixon, M.~Dusenberry, L.~Zhang, G.~Jerfel, and D.~Tran, ``Measuring
  calibration in deep learning,'' \emph{arXiv}, pp. arXiv--1904, 2019.

\bibitem{bauman2014dome}
S.~E. Bauman, T.~Benedict, M.~Baril, J.~C. Culiandre, I.~Look, G.~Matsushige,
  L.~Roberts, R.~Racine, D.~Salmon, and T.~Vermeulen, ``Dome venting: the path
  to thermal balance and superior image quality,'' in \emph{Observatory
  Operations: Strategies, Processes, and Systems V}, vol. 9149.\hskip 1em plus
  0.5em minus 0.4em\relax International Society for Optics and Photonics, 2014,
  p. 91491K.

\bibitem{salmon2009cfht}
D.~Salmon, J.-C. Cuillandre, G.~Barrick, J.~Thomas, K.~Ho, G.~Matsushige,
  T.~Benedict, and R.~Racine, ``Cfht image quality and the observing
  environment,'' \emph{Publications of the Astronomical Society of the
  Pacific}, vol. 121, no. 882, p. 905, 2009.

\bibitem{Pfrommer_2015}
\BIBentryALTinterwordspacing
T.~Pfrommer and P.~Hickson, ``High-resolution ground layer turbulence from
  inside the {CFHT} dome using a lunar scintillometer,'' \emph{Journal of
  Physics: Conference Series}, vol. 595, p. 012027, apr 2015. [Online].
  Available: \url{https://doi.org/10.1088%2F1742-6596%2F595%2F1%2F012027}
\BIBentrySTDinterwordspacing

\bibitem{lstm}
S.~Hochreiter and J.~Schmidhuber, ``Long short-term memory,'' \emph{Neural
  computation}, vol.~9, no.~8, pp. 1735--1780, 1997.

\bibitem{vae_missingvalues}
M.~Collier, A.~Nazabal, and C.~K. Williams, ``Vaes in the presence of missing
  data,'' \emph{arXiv preprint arXiv:2006.05301}, 2020.

\bibitem{imputation_missingvalues}
K.~Wo{\'z}nica and P.~Biecek, ``Does imputation matter? benchmark for real-life
  classification problems.'' \emph{ICML 2020 Workshop Artemiss}, 2020.

\end{thebibliography}
}

\end{document}